\documentstyle[12pt]{article}
\title{CHIRAL EFFECTS  IN THE CONFINING  QCD VACUUM}
\author{Yu.A.Simonov\\
Institute of Theoretical and Experimental Physics\\ 117259,Moscow
, B.Cheremushkinskaya 25, Russia \thanks{e-mail simonov at
vxitep.itep.msk.su}}
\date{}
\newcommand{\be}{\begin{equation}}
\newcommand{\ee}{\end{equation}}
\begin{document}
\maketitle

\begin{abstract}

Confining configurations are introduced into the standard instanton vacuum
model. This drastically improves theoretical properties of the vacuum:
instanton size density $d(\rho)$ stabilizes at $\rho\sim 0.2 fm$, all chiral
effects are formulated in a gauge-invariant way and
quarks are confined. An interesting  interplay of
chiral and confining dynamics is observed; for the
realistic values of parameters the Georgi-Manohar
picture emerges with chiral radius  $R_{ch}\sim
\rho\sim 0.2 fm$ much less than confining radius
$R_c\sim$ hadron radius $\sim 1 fm$.  In the limit
$R_{ch}\ll R_c$ the chiral mass $M_{ch}(p)$ is
unaffected by confinement and can be taken in the
local limit $M_{ch}(p=0)$.

Different types of effective chiral Lagrangians (ECL) are obtained,
containing all or a part of gluon, quark  and Nambu--Goldstone--meson
fields. The ECL are  manifestly gauge--invariant and in the limit of no
gluon fields coincide with those found previously.

 The problem of the double role of the pion -- as a
Goldstone meson or as a $q\bar{q}$ system is briefly
disscussed using confining ECL with quarks, mesons and
gluons.
\end{abstract}
\newpage

\section{Introduction}
\setcounter{equation}{0} \def\theequation{1.\arabic{equation}}

 Effects of chiral symmetry breaking (CSB) and in particular dynamics of
 Nambu-Goldstone bosons have been studied for a long time even before the
 discovery of QCD [1].

 The effective chiral Lagrangian (ECL) has been introduced [2] before the
 notion of quarks appeared in physics. Nowadays ECL and chiral
 perturbation theory is a powerful method [3] and only the loop cut-off
 problem and numerous phenomenological coefficients of ECL remind that this
 is an effective theory, not yet deducible from the first principles of QCD.

 What is the relation of confinement to chiral physics?
 At first sight two phenomena are not directly connected. This is
 explicitly so in the instanton model of chiral vacuum [4,5], where
 instantons do not bear confinement [6], but create zero modes [7] which
 are necessary for the quark condensate [8] and a fortiori
 for CSB [9,10]. At the same time there is a coincidence of chiral and
 deconfining phase transition temperatures in lattice calculations [11], and
 some attempts exist [12] to explain the fact in the framework of the
 instanton model.

 At present there are two disconnected approaches in quark physics, which
 treat confinement or chiral dynamics.
  The last approach uses ECL and chiral perturbation theory, where the
   notion of confinement is absent [2,3]. More microscopic among models of
 this type  is the instanton gas or liquid model [4,5]. This model yields
 microscopic chiral dynamics of quarks and produces ECL with coefficients
 which are calculable within the model. However quarks are deconfined and
 gauge invariance in the resulting ECL is violated. This of course is the
 price for keeping only instantons and disregarding confining degrees of
 freedom in the vacuum.  The first approach is exemplified by the so-called
 relativistic potential model (RPM) [13], where chiral effects are
 disregarded fully and all mesons including Nambu-Goldstone bosons are
 treated as bound state of $q\bar{q}$ interacting via linear potential.

 Recently a more general approach has been used for the $q\bar{q}$ system (for
a review see [14]), starting from QCD
 Lagrangian, where due to
 confinement a string was shown to appear for nonzero orbital momentum $L$,
 while at small $L$ one has RPM.  In this way in [15] RPM has
 been derived from QCD.

 But chiral as well as spin effects have been disregarded in [15].

 In this paper we start to consider  systematically chiral and confinement
 effects on the same footing. To this end we consider QCD vacuum  as
 consisting of instantons $A_{\mu}^{(i)}$ plus confining configurations
 $B_{\mu}$. Instantons create zero modes, which after being mixed and
 shifted due interaction give rise to CSB, chiral quark mass and ECL, as it
 was shown in the instanton liquid model in  [10,16].

 Confining background, superimposed on instantons, plays double role. First,
 and very important, it strongly modifies instanton density $d(\rho)$ at
 large instanton size $\rho$. This stems from the fact that  charge
 renormalization  at large distances in presence of $B_{\mu}$ strongly
 modifies [17]: logarithmic infrared divergence disappears, and the charge
 freezes at large distances [17]
 \be
 \alpha_s(R) \approx \frac{4\pi}{b_0ln\frac{R^{-2}+m^2}{(\Lambda_{QCD})^2}}
 \ee
 where $m$ is the lowest excitation mass of the order of 1 GeV.  As a
 consequence $g^2(\rho)$ in the renormalized instanton action $S_{inst}=
 \frac{8\pi^2}{g^2(\rho)}$ does not grow at large $\rho$ and the instanton
 density converges [18]
 \be
 \frac{d\rho}{\rho^5}d(\rho)= \frac{d\rho}{\rho^5} const
 [\frac{8\pi^2}{g^2(\rho)}]^{2N_c}e^{-\frac{8\pi^2}{g^2(\rho)}}
 \ee

 Thus confinement stabilizes the instanton size $\rho$ at some relatively
 small value $\rho \sim \rho_0$,
 \be
 \rho_0 \sim \frac{1}{m} \sim 1 GeV^{-1} \sim 0.2 fm
 \ee
 This value is slightly less than obtained in the instanton liquid model
 [5,10], where the size of instanton is fixed by instanton interaction at
 $\rho\approx 0.3 fm$. The important point here is that now $\rho_0$ depends
 on fundamental properties of the QCD vacuum rather than on
 model- dependent properties of instanton-instanton
 interaction.
 Second, confinement introduces important dynamical effects in calculation of
 all chiral effects, e.g. in calculation of the coefficients of
 ECL. In what follows we  write a new ECL modified by confinement
 which is manifestly gauge invariant. Here one can see
 explicitly how two dynamics, chiral and confining, interplay  and in
 particular one can check some provisions of Georgi and Manohar [19]. It
 appears that indeed the chiral scale $\Lambda_{CSB} \sim \rho^{-1}$ and the
 scale of confinement, coinciding with average size of a hadron
 $\Lambda_{QCD}$, are different and one may disentangle in some
 cases two dynamics. In particular the chiral mass in the limit
 $\Lambda_{CSB}\ll \Lambda_{QCD}$ becomes a local operator
 dependent only on chiral dynamics. On larger scale the chiral
 massive quark interacts with antiquark via a string and this
 creates a new quark mass --constituent quark mass.

 Thus the latter contains three ingredients:
 (i)current (or Lagrangian) mass,
 (ii) chiral mass,
 (iii) string mass. In  magnetic moments or spin-dependent forces exactly this
 constituent mass enters.

 Another interesting aspect of the interplay between chiral
 (instanton) and confining dynamics is the problem of double
 role of pion -- as a Nambu-Goldstone particle or as a bound
 state of quark and antiquark. We obtain an expression for the
 $q\bar{q}$ Green's function which contains both chiral and
 confining effects (i.e. the string due to nonperturbative
 gluons and pion exchanges) and is an extension of  our earlier
 result [14,20] where only confinement was taken into account, and of
 earlier result of [16, 21] where only chiral effects  have been considered.
 One can visualize there two possible type of poles, as
 suggested in [19] when one expands in pion field, and their
 position is separated (parametrically a Nambu-Goldstone pole
 at $m=0$  and $q\bar{q}$ pole at $m_a\sim \sqrt{\sigma}$).
  We show explicitly that the second type of pole at $m_a$ ( a "quark
 model pole") exactly cancels in the expression for the total Green's
 function, and only Nambu-Goldstone pole survives, while new  poles of
 unified dynamics appear heavier than $m_a$. This is in agreement with the
 scenario suggested by Georgi and Manohar [19].

   The  paper is organized as follows.

 In section 2 we study the quark propagation  in the instantonic vacuum in
 presence of confining background $B_{\mu}$. Methods of multiple scattering
 theory in $4d$ are widely used and the framework of the Dyakonov and Petrov
 approach [10] is modified to include $B_{\mu}$. In this modification all
 expressions are manifestly gauge covariant in contrast to the original ones
 of [10].

 Meanwhile we have justified in Appendix A the accuracy of the usually done
 approximation -- keeping only zero mode in the quark  Green's  function in
 the field of a given instanton: we show that omitted terms are of the order
 $m\rho\ll 1 $ as compared to the contribution of the zero mode. Another
 important estimate in Appendix B concerns the shift of the zero mode
 eigenvalue due to the presence of confining field $B_{\mu}$. It is shown
 that the shift $\delta \lambda$ is  $\leq 40  MeV$ and satisfies
 $\delta\lambda\cdot \rho \ll 1$, thus making all the method of [10]
 applicable also for the  realistic inclusion of confining field.

 In section 3 we are applying  another approach using a specific
  effective action suggested in [21-24] to calculate chiral  effects for
 any number of flavours. To
 this end a new simple derivation of the effective action is  given in
 Appendix C, which takes into account $B_{\mu}$. In the same section 3 we
 consider interplay of chiral and confining effects in the example   of the
 quark chiral mass and explicitly show that  these effects can be
 disentangled when their ranges are much different.

 The case of two flavours is considered in Section 4. Bosonization
method of [21] is slightly modified by confinement. One obtains here the
same "gap equation" for the chiral mass as for $N_f=1$.

In section 5 based on bosonization results of the previous section we deduce
ECL for quarks, pions and $B_{\mu}$ and also for pions and
$B_{\mu}$ only (after integration over quarks).

We also obtain here and in the Appendix D  an expression for $F_{\pi}$,
which is a modification of that of [10,21] due to confinement. In section 6
we consider the $q\bar{q}$ Green's function and discuss contribution to it
from confinement and Nambu-Goldstone modes. Here the  double face of pion is
discussed and some remarks on the OZI rule and OZI-violating mechanisms are
made.

The last section 7 is devoted to summary and an outlook.

\section{Quark propagation in the  instantonic vacuum with the  confining
background.}
\setcounter{equation}{0}
\def\theequation{2.\arabic{equation}}

In this Section we shall extend the method of [10] to the case  when  the
QCD vacuum contains in addition to the gas of instantons and antiinstantons
also some confining configurations, using mostly the method and notations of
ref. [10]. Our goal is to write the quark propagator and effective action in
the one-flavour  case in the form where the gauge invariance and confinement
are present explicitly.

We start with the ansatz for the vacuum
\be
A_{\mu}(x) = \sum^N_{i=1} A_{\mu}^{(i)}(x)+B_{\mu}(x)
\ee
where $N=N_++N_-$ is the total number of instantons and antiinstantons in
the  4-volume $V,A^{(i)}_{\mu}$ is  the field of (anti)instanton in the
singular gauge and $B_{\mu}(x)$ is the background  confining field which
ensures the observed string tension $\sigma$ through its field correlators
$<F_{\mu\nu}(x) F_{\lambda\sigma}(y)>$ etc. [25]. Note that the instanton
gas does not provide nonzero string tension,therefore the known value of
$\sigma \cong 0.2 GeV^2$  fixes  the normalization of the background field
$B_{\mu}$.

We define as in [12,26] the total quark Green's function $S$, "free" Green's
function $S_0$ and the quark propagator $S^{(i)}$ in the $i$-th
(anti)instanton  field:
\be
S=(-i\hat{D} (A)- im )^{-1}, ~~~
S^{(i)}=(-i\hat{D} (B+A^{(i)})- im )^{-1}, ~~~
\ee
$$S_0=(-i\hat{D} (B)- im )^{-1}, ~~~
$$

We also introduce a complete set of real eigenvalues $\lambda_n^{(i)}$ and
 eigenfunctions $u_n^{(i)}$ on a given center $i$:
 \be
 -i\hat{D}(B+A^{(i)}) u_n^{(i)}= \lambda_n^{(i)}u_n^{(i)},~~
 1\leq i\leq N; 0\leq n <\infty
 \ee
  So that $S^{(i)}$ can be written as
  \be
  S^{(i)}=\sum_n\frac{u_n^{(i)}(x)u_n^{(i)+}(y)}{\lambda_n^{(i)}-im}
  \ee
  with the definition
  \be
  t_i \equiv S_0-S^{(i)}
  \ee
  one can write the exact equations for the total quark propagator $S$ [26]
  \be
  S=S_0-\sum_{i,k} Q_{ik}~,
  \ee
  \be
  Q_{ik} = t_i\delta _{ik} - t_i S_0^{-1}\sum_{j\not=i }Q_{jk}
  \ee

  Eqs.(2.6)-2.7) are exact for the the given decomposition of the gauge
  field in (2.1). One can now make an approximation for $t_i$ using the fact
  that the sum over large values of $n$ in (2.4) is close to the "free"
  Green's  function $S_0(x,y)$, since $\lambda_n\sim \sqrt{p^2},~~ n\sim
  p_{\mu}$ at large $n$ and $p^2\gg <B^2_{\mu\nu}>$. Therefore one can
  approximate $t_i$ by a finite sum
  \be
  t_i (x,y) =-\sum_{n=0}^K
  \frac{u_n^{(i)}(x)u_n^{(i)+}(y)}{\lambda _n^{(i)}
  -im}
  \ee
  In what follows we shall often keep only the lowest term $n=0$,
  $\lambda_0^{(i)}\approx 0$, this approximation was exploited  in a series
  of papers [10,23,24]:
  \be
  S^{(i)} (x,y) = S_0 (x,y) +
  \frac{u_0^{(i)}(x)u_0^{(i)+}(y)}{\lambda _0^{(i)}
  -im}
  \ee
  We study the accuracy of the approximation (2.9) in Appendix A.
  We show there that the omitted terms are of the order of
  $0(m\rho)$ as compared to lowest mode contribution.  Another important
  topic concerns the shift of eigenvalues $\lambda_n^{(i)}$, defined in
  (2.3-2.4), due to the background field $B_{\mu}$. We  argue in Appendix B,
  that this shift is insignicant for the realistic values of gluonic
  condensate and instanton radius $\rho$. In particular,
  $(\lambda_0^{(i)})^2$ is of the order of $(30-40 MeV)^2$ and therefore
  does not spoil the approximation (2.9).

  Insertion of the separable form (2.8) into (2.7) yields the following
  solution for $S$:
  \be
  S(x,y) = S_0(x,y)- \sum_{i,k;n,m} u_n^{(i)} (x) (im - \hat{\lambda} +
  \tilde{V})^{-1}_{nm,ik} u_m^{(k)+}(y)
  \ee
  where we have defined the matrices $\lambda_{mn,ik}, V_{nm,ik}$
  \be
  \lambda_{nm,ik}= \delta_{ik}\delta_{nm} \lambda_{n}^{(i) }
  \ee
  \be
  \tilde{V}_{nm,ik}=\int u_n^{(i)+}(z)(-i\hat{D}(B) - im) u_m^{(k)} (z) d^4z
  \ee
  by definition $\tilde{V}_{nm},(i=k)\equiv 0$ .

  The form (2.10) coincides with that used in [10], when one keeps in the
  sum only $n=m=0 ,~~\lambda_0=0$.

  In  the rest of this section we shall calculate the  quark Green's
  function $S$ averaging (2.10) over vacuum configurations $B_{\mu}$ and
  (anti) instanton positions and orientations. We shall follow here the
  direct approach of [10], which makes explicit the appearing of the chiral
  mass of the quark, while in the next sections we follow another approach
  [21-24], where the notion of the effective action is introduced from the
  beginning. The new element which we shall obtain in this section is the
  gauge covariant form of the quark propagator with confinement taken into
  account.

  For the latter and for the averaging over $B_{\mu}$ and $A_{\mu}^{(i)}$
  one must define a physical gauge-invariant quantity associated with
  $S(x,y)$ (which itself is not gauge invariant).

  The simplest quantity is the Green's function of one light and one heavy
  quark in the limit when the heavy mass is infinitely large.
  \be
  G_{HL}(x,y)=<tr\Gamma S(x,y) \Gamma\Phi(y,x)>_{A_{\mu}^{(i)}, B_{\mu}}
  \ee
  Here $tr$ is a trace over color and Lorentz indices,
   $\Gamma = \gamma_5,\gamma_5\gamma_{\mu}, 1, \gamma_{\mu}$ etc., and
  \be
  \Phi(y,x)= P exp ~~ig \int^y_x B_{\mu}(z) dz_{\mu}
  \ee
  The integral in (2.14) is taken along the  straight line connecting $x$
  and $y$- this is the remnant of the heavy quark Green's function.

  The angular brackets in (2.13) denote averaging over fields $B_{\mu},
  A_{\mu}^{(i)}$ defined as follows.

  \be
  <0(A_{\mu}^{(i)}, B_{\mu})>_{A_{\mu}^{(i)}, B_{\mu}} \equiv
  \int \prod^N_{i=1}\frac{d^4R^{(i)}}{V} d\Omega_i d\mu(B)
  0(A_{\mu}^{(i)}, B_{\mu})
  \ee
  where $R_i$ is the position of (anti)instanton, $\Omega_i$ is its color
  orientation, and $d\mu(B)$ is the standard integration measure for the
  field $B_{\mu}$, the specific form of it is not needed for our purposes.

  The total gauge transformation for the field $A_{\mu}$
  \be
  A_{\mu}(x)\rightarrow U(x) (A_{\mu}(x) + \frac{i}{g} \partial_{\mu})U^+(x)
  \ee
  can be conveniently split into a homogeneous one for $A^{(i)}_{\mu}$ and
   inhomogeneous for $B_{\mu}$
   \be
   A_{\mu}^{(i)} \rightarrow U(x) A^{(i)}_{\mu}U^+(x)
   \ee
    \be
     B_{\mu}(x)
  \rightarrow U(x)(B_{\mu}(x)+\frac{i}{g} \partial_{\mu})U^+(x)
  \ee
  From(2.1) and (2.17-2.18) one returns back to (2.16). The color
  orientation for $A_{\mu}^{(i)}$ can be made explicit using
  \be
  A_{\mu}^{(i)} (x) = \Omega_i \bar{ A}_{\mu}^{(i)} \Omega^+_i
  \ee
   where
  $\bar{A}_{\mu}^{(i)}$ is the standard  singular gauge form [7,27]
  \be
  \bar{A}_{\mu}^{(i)}(x)= \frac{2}{g} \bar{\eta}_{a\mu\nu}(x-R^{(i)})_{\nu}
  \frac{\rho^2}{(x-R^{(i)})^2[(x-R^{(i)})^2 +\rho^2]}
  \ee

  It is clear from (2.17) that under global gauge transformation $\bar{U}$
  the constant  matrix $\Omega_i$ is simply "rotated"
  \be
  \Omega_i \rightarrow \bar{U} \Omega_i
  \ee
  We now turn to eigenfunction $u_n^{(i)}$. Under gauge transformation it
  transforms as
  \be
  u^{(i)}_n (x)\rightarrow U(x) u^{(i)}_n(x)
  \ee

  To make gauge dependence in $u_n^{(i)}$ explicit, we write it in the form
  \be
  u^{(i)}_n(x) = \Phi(x,R^{(i)}) \Omega_i\varphi_n(x-R^{(i)})
  \ee
  where
  \be
  \Phi(x,R^{(i)}) = P exp (ig \int^x_{R^{(i)}} B_{\mu}dz_{\mu}),
  \ee
  and $\varphi_n$ is the  form of solution in the singular gauge, e.g. for
  the (anti)instanton zero-mode solution one has [7,28]
  $$\varphi_0(x) = \bar{\varphi} (x) v^{\pm}_{\alpha m},$$
  \be
  \bar{\varphi}(x) =\frac{1}{\pi} \frac{\rho}{(x^2+\rho^2)^{3/2} }
  ~~~\frac{x_{\mu}\gamma_{\mu}}{\sqrt{x^2}}
  \ee
  with $v^{\pm}_{\alpha m} = \frac{1}{\sqrt{2}}\varepsilon^{\alpha
  m}(^1_{\pm 1})$ and $\alpha, m$ spin and color $SU(2)$ indices, and + and
  -- referring to the instanton and antiinstanton zero mode respectively.

  Since $\Phi(x, R^{(i)})$ transforms as
  \be
  \Phi (x, R^{(i)}) \rightarrow U(x) \Phi(x, R^{(i)}) U^+ (R^{(i)})
  \ee
  one can satisfy (2.22) imposing on $\Omega_i$ the transformation law
  consistent with (2.21)
  \be
  \Omega_i \rightarrow U(R^{(i)}) \Omega_i
  \ee
  It is clear now that $V_{nm, ik}$ (2.12) is gauge invariant, while
  $S(x,y)$ in (2.10) transforms in standard way:
  \be
  S(x,y) \rightarrow U(x) S(x,y) U^+(y)
  \ee
  and $G_{HL}$ (2.13) is gauge invariant.
  We do now a drastic approximation as in [10] to keep in (2.10) only terms
  with $n = m = 0$, and the lowest eigenvalue $\lambda_0$. In accordance
  with [10] we  define
  \be
  (\frac{1}{i\bar{m} +\hat{V}})_{ik}= \frac{\delta_{ik}}{i\bar{m}} +
  \left\{ \begin{array}{l}
  D_{ik}(R^{(i)}, R^{(k)}, \Omega_i, \Omega_k), i,k~~ \mbox{of one tipe}\\
  P_{ik}(R^{(i)}, R^{(k)}, \Omega_i, \Omega_k), i,k~~ \mbox{of diff. types}
  \end{array}
  \right.
  \ee
  where $i,k$ of one type  means both $i$ and $k$ are instantons or both
  antiinstantons and we have used notation $i\bar{m}=im-\lambda_0$.
  Following the same line of reasoning as in [10] we obtain the following
  equations for $D,P$ where we suppress arguments for simplicity
  \be
  P_{ik}= - \frac{1}{i\bar{m}} V_{ik} \frac{1}{i\bar{m}} - \frac{N}{2V}
  \int d^4R^{(i)}d\Omega_j \frac{1}{i\bar{m}} V_{ij} \times
  \frac{1}{1-i\bar{m}\delta}D_{jk}
  \ee
  \be
  D_{ik} = -\frac{N}{2V}\int d^4R^{(j)} d\Omega_j \frac{1}{i\bar{m}}V_{ij}
  \frac{1}{1-i\bar{m}\delta} P_{jk}
  \ee
  Several comments are in order. First, $P_{ij}$ and $D_{ij}$
can be  considered as probability amplitudes for  a quark to travel from a
center $i$ to a center $j$ with all  possible centers being on its way.
Both $P_{ij}$ and $D_{ij}$ are defined not to contain (infinitely many, in
principle) returns to the centers $i,j$, and those should be accounted for
separately (note that (2.29) accordingly is a precise  definition only for
paths without returns).

This is done introducing the factor
\be
(1-i\bar{m}\delta)^{-1} = \sum^{\infty}_{n=0} (i\bar{m}\delta)^n
\ee
where $i\bar{m}\delta$ is the amplitude of returning to the center only
once,
\be
 \delta\equiv D_{ii} (R^{(i)}, R^{(i)}, \Omega_i, \Omega_i)
 \ee
Second, averaging over all $R^{(j)}, \Omega_j$ with $j\not= i,k$ is assumed
in $D_{ik}, P_{ik}$. This can be factorized out only in the limit $N_c
\rightarrow \infty$ and this limit is assumed everywhere below.

It is convenient to factorize explicitly the dependence on $\Omega $ in
$D_{ik}, P_{ik}$
\be
D_{ik} = v^+_{\alpha_i m} (\Omega^+_i)_{m \nu}
d^{ik}_{\nu\nu'}(\Omega_k)_{\nu'\rho}v_{\rho\beta_j}
\ee
\be
P_{ik}=  v^+_{\alpha_i m} (\Omega^+_i)_{m \nu}
f^{ik}_{\nu\nu'}\Omega v
\ee
\be
V_{ik}= v^+_{\alpha_i m} (\Omega^+_i)_{m \nu}
V^{ik}_{\nu\nu'}\Omega v
 \ee
{}From (2.34) one concludes that the $\delta $ does not depend on $i$ in
(2.33) and is a gauge-invariant quantity $(N_c\rightarrow \infty)$, if one
exploits the fact that it is diagonal in spin indices $ \alpha,
\beta(cf(2.25))$
\begin{eqnarray}
 \delta &=& \frac{1}{2}
Tr_{\alpha}(D_{ii}(R^{(i)}, R^{(i)}, \Omega_i, \Omega_i)) = \\ \nonumber &=&
\frac{1}{2} Tr_{\alpha} v^{+}_{\alpha m}(\Omega_i)_{m \nu} d^{ii}_{\nu
\nu'}(\Omega_i)_{\nu' \rho} v_{\rho \alpha} = \\
\nonumber
&=& \frac{1}{4} Tr_{\nu} d_{\nu\nu}^{ii}
\end{eqnarray}
The color trace appearing in the last equality in (2.37) signals the gauge
invariance of $\delta$.
We are interested in the limit $m \rightarrow 0$, therefore we introduce the
reduced quantities $\bar{d}, \bar{f}, \bar{v}$ where dependence on $\bar{m}$
and $\frac{N}{V}$ is made explicit,
\begin{eqnarray}
\nonumber
d^{ik} = \frac{2VN_c}{N{\varepsilon}} \frac{i}{\bar{m}^2} \bar
{d}^{ik}~,~~
f^{ik} = \frac{2VN_c}{N\varepsilon} \frac{i}{\bar{m}^2}\bar{f}^{ik} \\
V^{ik} = \frac{2VN_c}{N\varepsilon i} \bar{V}^{ik}~ ;~~
 \varepsilon \equiv \frac{1}
{\bar{m}(1-i\bar{m}\delta)}
\end{eqnarray}
and obtain equations not containing $m$ any more
\begin{eqnarray}
 \bar{f}^{ik}=-\bar{V}^{ik} +\int \bar{V}^{ij} dR^{(j)}\bar{d}^{jk}\\
\nonumber \bar{d}^{ik} = \int\bar{V}^{ij} dR^{(j)} \bar{f}^{jk}
\end{eqnarray}

The fundamental role in the chiral mass generation is played by the
so-called "consistency equation" [10] which in our notations is
\be
Sp_c \bar{d}^{ii} = \frac{2N}{VN_c}
\ee
Using (2.39) it can also be written as
\begin{eqnarray}
\frac{1}{\varepsilon} &=& \frac{1}{4}  \{ \frac{N}{2VN_c} \varepsilon
Sp_c(V_{ik} V_{ki}) + \\
\nonumber
&+& 0(\frac{N}{2VN_c})^2 \varepsilon^2 tr(VVV)
\}
\end{eqnarray}
The first term in (2.41) can be written  explicitly:
\begin{eqnarray}
Sp_c(V_{ik}V_{ki}) \equiv  Sp_c \int d^4z d^4 z' d^4
R^{(k)}\bar{\varphi}^+(z - R^{(i)}) \Phi(R^{(i)}Z) i\hat{D} \Phi(z,R^{(k)}
 \times \\
\nonumber
 \times \bar{\varphi}(z - R^{(k)}) \cdot \bar{\varphi}^{+}(z'-R_k)
\Phi(R^{(k)},z') i\hat{D} \Phi (z', R^{(i)}) \bar{\varphi}(z'-R{(i)})
\end{eqnarray}
We note that $d^{ik}, f^{ik}$ and $V^{ik}$ as well as
$\bar{d}^{ik},~~\bar{f}^{ik},~~\bar{V}^{ik}$ are gauge covariant and
transform as
\be
(f^{ik}, V^{ik}, d^{ik}) \rightarrow U(R^{(i)}) (f^{ik},V^{ik},d^{ik})
U^+(R^{(k)})
\ee
In contrast to that $\delta$ and $\varepsilon$ are gauge invariant, as can
seen in (2.37) and (2.42).

The measurable physical quantity associated with $f$, $d$ is the heavy-light
Green's function $G_{HL}$ and we use (2.10), (2.13), (2.29) to write it in
the form similar to that of Eq.(40) from [10]  (we remind the reader that
$D^{ik}$ and $F^{ik}$ are defined without returns to the centers $i$ and $k$
and therefore the returns, i.e. factors like $\frac{1}{1-im\delta}$, should
be inserted in (2.10)).
$$<G_{HL}(x,y)>_{B,A^{(i)}}= <\Phi(y,x) \Gamma S_0(x,y)\Gamma>- $$
$$-\frac{N}{2VN_c}(\frac{1}{i\bar{m}}+\frac{\delta}{1-i\bar{m}\delta})$$
$$\times <\Gamma\Phi(y,x)\Gamma \bar{\varphi}(x-R_i) \Phi(x, R_i) \Phi(R_i,
y)\bar{\varphi}(y-R_i)>_{B,R_i} $$
\be
-(\frac{N}{2VN_c})^2(\frac{1}{1-i\bar{m}\delta})^2
\ee
$$\times <\Gamma\Phi(y,x)\Gamma
 \bar{\varphi}(x-R_i) \Phi(x, R_i) d_{ik}(R_i,
R_k)\Phi(R_k,y)\bar{\varphi} (y-R_k)>_{B,R_i,R_k}
$$
$$+ (inst.\rightarrow antiinst.)$$
$$-(\frac{N}{2VN_c})^2(\frac{1}{1-i\bar{m}\delta})^2$$
$$\times <\Gamma\Phi(y,x)\Gamma
 \bar{\varphi}(x-R_i) \Phi(x, R_i) f_{ik}(R_i,
R_k)\Phi(R_k,y)\bar{\varphi} (y-R_k)>_{B,R_{ij},R_k}$$
$$-(inst.\rightarrow antiinst.)$$

When the field $B_{\mu}$ is put equal to zero, one comes back to the
situation studied in [10]. In this case it is useful to work in the momentum
space and one obtains as in [10]
\be
S(p)\sim <G_{HL}(p)> = \frac{i\hat{p}+M(p)}{p^2+M^2(p)}
\ee

with
\be
M(p) = \frac{\varepsilon N}{2VN_c} p^2\bar{\varphi}^2(p)
\ee
and the equation for $\varepsilon$ ("consistency equation", which one can
call also the "gap equation") obtains from (2.40) putting all $\Phi=1$, \be
\frac{4VN_c}{N}\int\frac{d^4p}{(2\pi)^4}\frac{M^2(p)}{M^2(p)+p^2} =1
\ee
This is the equation (36) of ref. [10] in the limit $m\rightarrow 0$.

The corresponding partition function can be written as [21]
\be
Z_{QCD}(N_f=1)= const \int D\psi D\psi^+ exp [-\int \frac{d^4p}{(2\pi)^4}
\psi^+(p)(\hat{p}-iM(p))\psi(p)]
\ee
Note that the resulting theory of massive quarks with the chiral mass $M(p)$
is not gauge invariant, as is clearly seen in (2.48). This serious defect is
cured in (2.44) which is fully gauge invariant, but  the notion of the
quark Green's function $S(p)$ and the chiral mass can not be made gauge
invariant in contrast to $G_{HL}$.

\section{Method of effective action -the case of one flavour}
\setcounter{equation}{0}
\def\theequation{3.\arabic{equation}}

In this section we adopt another procedure -- the method of effective action
suggested in [21-24]. One can prove [24] that this method yields the same
Green's function as in the more direct but tedious method of [10], which we
have used in the previous section.

The proof given in [24] contains intermediate formulas which are diverging
for $m\rightarrow 0$, i.e. the partition function $Z \sim m^{-NN_f}$.
 Therefore we give a different derivation in Appendix C and
check at each point  the accuracy of physical approximations. In particular
we show in Appendix C that there is no divergency for $m\rightarrow 0$ and
that  leading term in the limit $m\rho \ll 1$ is the same as obtained in
[24] when $B_{\mu}= 0$.

As shown in Appendix C the QCD partition function for quarks in the field
(2.1) can be written as (in the limit $m\rightarrow 0$)
\begin{eqnarray}
Z_{QCD}= \int D\psi D\psi^+ D\mu(B) e^{\int d^4x
\psi^+_fi\hat{D}(B)\psi_f}\times\nonumber \\
\times <\prod^{N_f}_{f=1}\prod^{N}_{i=1}(2im_f-\int d^4x \psi^+_f i\hat{D}
u_0^{(i)}(x) \int d^4yu_0^{(i)+}(y) i\hat{D}\psi_f(y))>_{R_i,\Omega_i}
\end{eqnarray}
Here $f$ refers to the flavour (total number of flavours is $N_f$)
the meaning of $2im_f$ is explained in Appendix C eq. C.8, and
$u_0^{(i)}(x)$ is defined in (2,23) and we did not specify the difference
between instantonic and antiinstantonic zero modes, which should be kept in
mind in (3.1).

We now consider systematically the cases of $N_f=1$ in this section and the
case of $N_f=2$ in the next section.\\

\underline{The case of $N_f=1$.}

In this case the averaging procedure over $R_i,\Omega_i$ factorizes in (3.1)
and one can introduce a two-fermion vertex (to stress the analogy with [21]
we keep notations of that paper)
\begin{eqnarray}
Y_{\pm} = \int dR_i d\Omega_i \int \psi^+(x)i\hat{D}u_0^{(i)}(x)d^4x \int
d^y u_0^{(i)^+}(y) i\hat{D}\psi(y)=\nonumber \\
= \int dR_i \psi^+(x)i\hat{D}\bar{\varphi} (x-R_i)\frac{1}{2}(1_{\mp}
\gamma_5)
\Phi(x,R_i,y) \bar{\varphi}^+(y-R_i) i\hat{D}\psi(y)dydx
\end{eqnarray}
where
$\Phi(x,R_i,y)\equiv \Phi(x,R_i) \Phi(R_i,y)$

Following [21] we introduce identically integrations over
$\lambda_+,\lambda_-, \Gamma_+, \Gamma_-$ to obtain
\be
Z_{QCD} = \int D{\mu}(B) D{\psi} D{\psi}^+\int^{\infty}_{-\infty}
\frac{d\lambda_+}{2\pi}
\int^{\infty}_{-\infty}
\frac{d\lambda_-}{2\pi}
\int^{\infty}_{-\infty}
d\Gamma_+
\int^{\infty}_{-\infty}
d\Gamma_-
exp W
\ee
$$ W= \int d^4x \psi^+ i\hat{D}\psi + i \lambda_+ (Y_+-\Gamma_+) +
i\lambda_-(Y_--\Gamma_-)+
N_+\ln\frac{\Gamma_+}{V}+N_-\ln\frac{\Gamma_-}{V}$$

Integrating over $\Gamma_+, \Gamma_-$ by the steepest descent method in the
thermodynamical limit $N_{\pm}\rightarrow \infty, V\rightarrow \infty,
\frac{N_{\pm}}{V}= const.$, one obtains
\be
Z_{QCD} = \int D{\mu}(B)
\frac{d\lambda_+}{2\pi}\frac{d\lambda}{2\pi}D{\psi}
D{\psi}^+
exp \bar{W}
\ee
$$\bar{ W}= N_+\ln \frac{N_+}{i\lambda_+V}-N_+
+N_-\ln\frac{N_-}{i\lambda_-V}-N_- + \int d^4x \psi^+ i\hat{D}\psi + i
\lambda_+
Y_+ + i\lambda_-Y_-$$

The integation over $d\lambda_+d\lambda_-$ can be also done by the steepest
descent method when $N_{\pm}$ is large [21].

Before doing that we integrate over $\psi,\psi^+$; using notation
\be
Y_{\pm}= \psi^+\frac{1\mp \gamma_5}{2} \bar{Y}_{\pm}\cdot \psi
\ee
we obtain
\be
Z_{QCD} = \int D{\mu}(B)
\frac{d\lambda_+}{2\pi}\frac{d\lambda_-}{2\pi}exp (W'+W"),
\ee
\be
 W'= N_+\ln \frac{N_+}{i\lambda_+V}-N_+
+N_-\ln\frac{N_-}{i\lambda_-V}-N_-
\ee
\be
W"= \ln det \left( \begin{array}{ll}
i\hat{D}&i\bar{Y}_+\lambda_+\\
i\bar{Y}_-\lambda_-& i\hat{D}
\end{array}\right) =
\ee
$$Tr \ln (-\hat{D}^2+ \bar{Y}_+\bar{Y}_- \lambda_+\lambda_-),$$
where $Tr$ means trace over coordinates, color, and Lorentz indices.
Integrating now over $d\lambda_+d\lambda_-$ one finds $\lambda_+,\lambda_-$
in the extremum of $(W'+W")$ to satisfy equations
\be
\frac{N_+}{\lambda_+} =
Tr\{\bar{Y}_+\bar{Y}_-\lambda_-(-\hat{D}^2+\bar{Y}_+\bar{Y}_-\lambda_+\lambda_-)^{-1}\}
\ee
$$\frac{N_-}{\lambda_-} =
Tr\{\bar{Y}_+\bar{Y}_-\lambda_+(-\hat{D}^2+\bar{Y}_+\bar{Y}_-\lambda_+\lambda_-)^{-1}\}$$
For $N_+=N_-=\frac{N}{2}$, one finds $\lambda_+=\lambda_-=\frac{\varepsilon
N}{2VN_c}$, where $\varepsilon$ satisfies the "gap equation"
\be
\frac{N}{2} = (\frac{\varepsilon N}{2VN_c})^2
Tr(\frac{\bar{Y}_+\bar{Y}_-}{-\hat{D}^2+\bar{Y}_+\bar{Y}_-(\frac{\varepsilon
N}{2VN_c})^2})
\ee

 Since according to (3.2), (3.5), one has for $\bar{Y}_{\pm}$,
 \be
 \bar{Y}_{\pm}=\int dR_i i\hat{D} \bar{\varphi}(x-R_i)
 \Phi(x,R_i,y) \bar{\varphi} (y-R_i) i{\hat{D}}
 \ee
 the r.h.s. of (3.10) and hence $\varepsilon$ is gauge invariant (only
 closed  "contours" of $\Phi(x,R_i,y)$ enter under                 the sign
 of $Tr$).

 In case when $B_{\mu}=0,~~ \Phi=1,~~ \hat{D}\rightarrow \hat{\partial}$,
 (3.11) reduces to
\be
Y_{\pm}(x,y) = \int \frac{dp}{(2\pi)^4}
e^{ip(x-y)}(\hat{p}\bar{\varphi}_{\pm}(p))^2
\ee
and (3.10) becomes the same as (2.47)
\be
\int \frac{d^4p}{(2\pi)^4}\frac{M^2(p)}{p^2+M^2(p)}\frac{4VN_c}{N}=1
\ee
with $M(p)$ given by (2.46).

Now back to the case $B_{\mu}\not=0$. We integrate in (3.4) over
$d\lambda_+d\lambda_-$ first, replacing in  $\bar{W}$ the $\lambda_{\pm}$ by
their extremal values (3.9) We obtain the effective action for quarks in the
form $(N_f=1)$
\be
Z_{QCD} = const \int D{\mu}(B)
D{\psi}
D{\psi}^+
exp \int dx dy \psi^+(x) [i\hat{D}\cdot \delta(x-y) +iM(x,y)]\psi(y)
\ee
where the nonlocal mass operator is
 \be
M(x,y) =\frac{\varepsilon N}{2VN_c} \int dR_i i\hat{D}
 \bar{\varphi} (x-R_i) \Phi(x,R_i,y) \bar{\varphi}^+ (y-R_i)
 i{\hat{D}}
 \ee
 and $\varepsilon$ is to be defined from (3.10).

 Note, that the effective action in (3.14) is now gauge invariant in
 contrast to (2.48), since from (3.15) one deduces that under gauge
 transformations $M(x,y)$ changes as
 \be
 M(x,y) \rightarrow U(x) M(x,y) U^+(y)
 \ee
 $M(x,y)$ as given by (3.15) is what one may call the chiral mass operator.
 Inclusion of background field  $B_{\mu}$ makes it gauge covariant, but now
 $M(x,y)$ is dependent on the confining forces, more explicitly, $ M(x,y)$
 contains the field $B_{\mu}$ and therefore depends on the string which
 connects quark with antiquark (or with the string junction  in baryon).

 To study this dependence more explicitly consider again the heavy-light
 Green's function, calculated with the help of (3.14):
 \be
 <G_{HL}(x,y)>_B = \int D\psi D\psi^+ D\mu(B) \psi(y) \Phi(y,x)\psi^+(x)
 e^{\psi^+(\hat{D}+\hat{M}) \psi}=
 \ee
 $$=tr_c(\Phi(y,x)(i\hat{D}+i\hat{M})^{-1}_{xy})$$
 We have neglected additional $q\bar{q}$ pairs (quenched approximation or
 large $N_c$ limit). Using the Feynman-Schwinger representation [14] (3.17
 can be rewritten as
 \be
 <G_{HL}(x,y)>_B =<tr_c(i\hat{D}-i\hat{M})
 \int^{\infty}_{0}ds Dz e^{-K} \psi_z(y,x)\Phi(y,x)>_B
 \ee
  where
  \be
 \psi_z(x,y) = P(e^{-s(\hat{M}^2-\Sigma F)}\Phi_z(x,y))~, ~~K =
 \int^s_0\frac{\dot{z}^2(\lambda)}{4}d\lambda~, ~~\Sigma F = \frac{1}{4}
 \sigma_{\mu\nu}F_{\mu\nu}
 \ee
  and the ordering operator $P$ ensures the proper insertion of the mass
  operator $\hat{M}^2$ into the phase factor $\Phi_z(x,y)$ where the
  subscript $z$ refers to the contour of $\Phi_z$ taken along the quark
  path.

  In (3.18) the field $B_{\mu}$ enters $\hat{M}$ and the closed contour $C$
  formed by the straight line and the quark path between $x$ and $y$, as
  shown in Fig.1, together with insertions of the mass operator.

  The dynamics in (3.18) is defined by the Wilson loop with mass insertions,
  shown in Fig. 1, which obeys the area law:
  \be
  <\psi_z(x,y) \Phi(y,x) > \equiv <W_M(C)> \sim exp (-S_M\cdot \sigma)
  \ee
  where $S_M$ is the minimal area of the contour $C$ with mass insertions.
Eq. (3.20) means that a string is formed between light and heavy quark
trajectories and this string is influenced by the insertions of the mass
operator.

To
understand better
 these insertions  one may look at the quark propagator and
expand it in powers of the  mass operator:
\be
(i\hat{D} +i\hat{M})^{-1}_{xy} =
(i\hat{D})^{-1}_{xy} -
(i\hat{D})^{-1}_{xu}d^4u
 i\hat{M}(u,v)d^4v
 (i\hat{D})^{-1}_{xy}
 \ee
 $$
 + (i\hat{D})^{-1}_{xu}d^4uiM(u,v)d^4v
 (i\hat{D})^{-1}_{vt}d^4tiM(t,w)d^4w
 (i\hat{D})^{-1}_{wy}-...
 $$

 The nonlocality of $M(x,y)$ as can be seen in (3.15) is of the order of the
 $\rho$ -- average size of (anti)instantons (the integrand in (3.15) behaves
 as $\frac{1}{|x-R_i|^4}\frac{1}{|y-R_i|^4}$ at $(|x-R_i|, |y-R_i| \gg
 \rho)$. At the same time the average size of a hadron in (3.18) - the
 width of the contour $C$ in Fig. 1 -- is of the order of the confinement
 radius $R_c\sim 0.5 fm$ for lowest states or larger than $R_c$ for
 excited states.

 Therefore we  have to distinguish two cases.\\
 i)$\rho\geq R_c$. In this case the nonlocality of $\hat{M}$ is strongly
 influenced and interrelated with the dynamics of the string -- one cannot
 separate effects of chiral symmetry braking (the chiral mass $\hat{M}$) and
 confinement.\\
 ii)$\rho\ll R_c$. In this case we can effectively replace in (3.15)
 $\Phi(x,R_i,y)$ by $\Phi(x,y)$ and hence rewrite $\hat{M}$ as
 \be
 M(x,y)=\Phi(x,y) \int M(p) e^{ip(x-y)}\frac{d^4p}{(2\pi)^4}
 \ee
 Furthermore we can neglect nonlocality of $M(x,y)$ on the scale of $(x-y)
 \sim R_c$ and replace the operator $M(x,y)$ by
 \be
 M(x,y)\approx \delta(x-y)\int d (x-y)
 M(x,y)=\delta(x-y) M(0)
 \ee
 Introduction of (3.23) into (3.14) yields now a gauge-invariant expression
 with constant chiral mass and eq.(3.18) assumes the form
  \be
 <G_{HL}(x,y)> =tr(i\hat{D}-iM(0)) \int ds Dz e^{-K-sM^2(0)} e^{-\sigma S}
 \ee
 where the effects of CSB and confinement are separated. The first produced
 the chiral mass $M(0)$ which enters instead of current mass, and the latter
 ensures the string dynamics between the now massive light quark and heavy
 antiquark. We conclude this Section with discussion of the quark
 condensate.  From (3.14) we have
 \be
  <\bar{q}q>_M =
 itr<(-i\hat{D}-i\hat{M})_{xx}^{-1}>_B
 \ee
  In the case $B_{\mu}=0$ (3.25)
 reduces to
 \be
  <\bar{q}q>_M=-\int
 \frac{d^4p}{(2\pi)^4}\frac{M(p)}{p^2+M^2(p)}
 \ee
  where $M(p)$ is the
 Fourier transform of (3.15) when $\Phi\equiv 1$, and is the same as in
 (2.46). We note that the form (3.26) coincides with  that given in [10,21].

 When $B_{\mu}\not= 0 $ one should calculate the original expression (3.25)
 which can be rewritten with the help of (3.24) as
 \be
 <\bar{q}q>_M=-<\int^s_0 tr \hat{M}Dz
 e^{-K}\psi_C(x,x)>_B
 \ee
 where the contour $C$ in $\psi_C$ is a closed trajectory of the quark, to
 be integrated over in $Dz$. From (3.26) one can notice that the integral is
 effectively defined by the distances of the order of the radius of
 instanton $\rho$. When this radius is assumed to be much smaller than the
 confinement radius $R_c$ (the case ii) above ), then the effects of
 confinement are unimportant and give only a small correction to (3.26).
 Thus we have two different situations: in computing $q\bar{q}$ system (wave
 functions and masses) one can use the inequality $\rho \ll R_c$ and keep
 only $M(0)$ instead of $M(p)$ in the first approximation in Green's
 function; in computing $<\bar{q}q>$ on the other hand one should keep
 dependence $M(p)$ and can neglect in the first approximation effects of
 confinement.

\section{The case of two flavours}
\setcounter{equation}{0}
\def\theequation{4.\arabic{equation}}

In the general case of $N_f$ flavours in the effective action in (3.1) there
enters a $2N_f$ vertex [21,24]
\be
Y_{\pm}=(-)^{N_f}\int d^4R_i d\Omega_i\prod^N_{f=1}\int d^4x\psi^+_f
i\hat{D}u^{(i)}\int d^4y u^{(i)}(y) i\hat{D}\psi_f(y)
\ee

Integrating over $d\Omega_i$ for $N_f=2$ and taking the limit
$N_c\rightarrow \infty$ one obtains
\be
Y_{\pm}=\int d^4R detJ_{\pm}(R)
\ee
where
\be
(J_{\pm}(R))_{fg}=\int dx dy \psi^+_f(x)
\frac{1}{2}(1\mp\gamma_5) K(x,y,R)\psi_g(y)
\ee
and
\be
K(x,y,R)=i\hat{D}\bar{\varphi}(x-R)\Phi(x,R,y)
\bar{\varphi}^+(y-R)i{\hat{D}}
\ee
The interaction (4.2) is reminiscent of the 'tHooft determinantal
interaction [7], and can be also compared to the similar term deduced in
[28].

The difference in our case, as also in the $B_{\mu}=0$ case of [21,28], is
that our interaction is nonlocal with nonlocality of the order of $\rho$ --
size of instanton. In addition, in contrast to [21,24] the term (4.2) is
gauge invariant and takes into account effects of confinement.

The effective partition function similarly to (3.4) is given by (up to
unessential factors and replacing $i\lambda_{\pm} \rightarrow g_{\pm}~,~~
R\rightarrow u$
$$
Z_{QCD} \sim \int dg_+\int dg_- D\psi D\psi^+ exp W_2
$$
\be
W_2 = \int \{\sum^2_{f=1}\psi^+_f i\hat{D}\psi_f(u)+g_+det
J_+(u)+g_-detJ_-(u)\} du
\ee
$$-N_+lng_+-N_-lng_-$$

One can introduce as in [21] the $2\times 2$ flavour matrices $\tau_a^-$,
$\tau^-_a=(\vec{\tau},i), a=1,...4,$ and use the identity
\be
(\tau^-_a)_{fg}(\tau^-_a)_{f\prime g\prime}
=-2\varepsilon_{fg\prime}\varepsilon_{gf\prime}
\ee
One can exploit the Hubbard-Stratonovich tranformation
and write $Z_{QCD}$ through
additional functions $L_a, R_a$ [21] to obtain
\be
 Z_{QCD}= \int D{\mu}(B)
dg_+ \int dg_- \int D\psi D\psi^+ DL_a DR_a exp \bar{W}_2
\ee
 $$\bar{W}_2 =
- N_+ lng_+ - N_-lng_-+\int d^4x [\psi^+_fi\hat{D}\psi_f+$$
$$+g_+L_a^2(x)+g_-R^2_a(x)+2g_+L_a(x)J_+^a(x)+$$
$$+2g_-R_a(x)J_-^a(x)]$$
where
\be
J_{\pm}^a(x)= (\tau_a^-)_{fg}(J_{\pm}(x))_{fg}
\ee

The effective action $\bar{W}_2$ contains both fermionic degrees of freedom
$\psi, \psi^+$ and bosonic $L_a, R_a$, and global parameters $g_+, g_-$.
Integrating over $D\psi D\psi^+$ we get a fully bosonic effective action
\be
e^{-W(L,R)}= \int dg_+ dg_- e^{-N_+lng_+-N_-lng_-+\int
d^4y(g_+L^2_a(y)+g_-R^2_a(y))} exp X;
\ee
\be
X= ln~ det\left( \begin{array}{ll}
i\hat{D}&2g_+\hat{L}K\\
 2g_-\hat{R}K&
i\hat{D}
\end{array}
\right)\\ ,
\ee
where $\hat{L}=L_a\tau_a^-,~~\hat{R}=R_a\tau_a^-$, and  e.g.$(\hat{L}K)_{xy}
 =\int\hat{L}(u)K(x,y,u)du$.

We  shall now study following [21] the phenomenon of CSB using variables
$L_a,R_a$. We shall look for the condensate of $L_4, R_4$, namely
introducing
\be
\sigma(x) = L_4(x) + R_4(x),
\ee
we show that
\be
<\sigma> \not= 0 , <L_4-R_4>=<L_i>=<R_i>=0
\ee
Since $\hat{L}$ and $\hat{R}$ transform under $SU_L(N_f)\times SU_R(N_f)$ as
\be
\hat{L}\rightarrow U_L\cdot \hat{L}U^+_R, ~~\hat{R}\rightarrow
U_R\hat{R}U^+_L \ee the nonzero value of $<\sigma>$ signals CSB.

 The insertion of (4.11) , (4.12) into (4.10) yields
\be
X= Tr ~ln(-\hat{D}^2+g_+g_-\sigma^2 K^2)
\ee

The integration over $dg_+dg_-d\sigma$ via the steepest descent method yields
equations for determination of $g^{(0)}_+, g^{(0)}_-, \sigma^{(0)}$
\be
\frac{\partial W}{\partial\sigma}= -\frac
{g_++g_-}{2}V\sigma - 2
\sigma g_+g_-Tr(\frac{K^2}{-\hat{D}^2+g_+g_-\sigma^2K^2})=0
\ee
$$
\frac{\partial W}{\partial g_{\pm}}=
+\frac{N_{\pm}}{g_{\pm}}-\frac{V\sigma^2}{4} -
\sigma^2 g_{\mp}Tr(\frac{K^2}{-\hat{D}^2+g_+g_-\sigma^2K^2})=0
$$

{}From (4.15) we find that $g_0=-\frac{N}{V\sigma^2_0}$, and
\be
g_0^2\sigma_0^2 Tr(\frac{K^2}{-\hat{D}^2+g_0^2\sigma^2_0
K^2})=+\frac{N}{2}
\ee

Comparing with the gap equation (3.10) we see that  $\sigma_0 =
\frac{2N_c}{\varepsilon}$ (our normalization differs from that of [21], note
also misprints in numerical coefficients in  Eqs. (27-29) of [21]).

We can now again define the quark effective action if we insert in $W_2$ in
(4.7) the extremal values of $L_a, R_a, g_+, g_-$ found above.

We obtain as before the effective action (3.14) when we express $\sigma_0 $
and $g_0$ in terms of $\varepsilon$.

Thus the one-quark situation in case of $N_f=2$ is the same as in the case
$N_f=1$: the effect of CSB is to create the chiral quark mass.

This is true however only in the approximation (4.12) when all the effects
of bosonization reduce to the  creation of nonzero $\sigma_0$, and no boson
exchanges (quantum boson fields $L_a, R_a$) are allowed. In the next section
we shall discuss these effects and derive effective chiral Lagrangian with
confinement taken into account.

\section{Effective chiral Lagrangian and quarks}
\setcounter{equation}{0}
\def\theequation{5.\arabic{equation}}

We are now in position to calculate the "bosonic" effective Lagrangian
$W(L,R)$ in (4.9), inserting there extremal values of $g_+=g_-=g_0$ and
$<\sigma> = \sigma_0$,\\
namely
\be
 g_0\sigma_0 = -\frac{\varepsilon N}{2VN_c}~,~~ \sigma_0 =
\frac{2N_c}{\varepsilon}
\ee
To parametrize the quantum bosonic fields in $\hat{L}, \hat{R}$ we introduce
as in [21] the forms
\be
\hat{L} = i\sigma_0 \frac{1+\sigma+\eta}{2} UV~,~~ \hat{R} = i\sigma_0
\frac{1+\sigma-\eta}{2} VU^+
\ee
where $U=exp(i\pi_i\tau_i)~,~~V=exp(i\sigma_i\tau_i)~,i = 1,2,3$.
The eight bosonic fields $\pi_i,\sigma_i,\sigma, \eta$ correspond to eight
field $L_a, R_a,~~ a=1,...4.$

Insertion of (5.2) into (4.9) yields
\be
W(L,R)=\frac{N}{2V}\int d^4x(\sigma^2(x)+\eta^2(x))-
\ee
$$Tr~ ln
\{i\hat{D}+i(1+\sigma+\eta)UV\hat{M}_++i(1+\sigma-\eta) VU^+\hat{M}_-\}$$
where $\hat{M}_{\pm}= \hat{M}\frac{1\mp\gamma_5}{2}$ and $\hat{M}
$ is defined in (3.15) and $Tr$ is taken over
coordinates, color and Lorentz indices. One should keep in mind that terms
linear in fields should be suppressed since they vanish due to the steepest
descent condition, yielding $<\eta>=<\pi_i> =<\sigma_i>=0$.

It is instructive to  expand the effective action (5.3) in $\pi_i,
\sigma_i,\sigma, \eta$  and to find the corresponding quadratic terms
yielding masses of mesons.

For pions this procedure looks like
$$-W(\pi)=Tr~ln(i\hat{D}+iU\hat{M}_++iU^+\hat{M}_-)=$$
\be
Tr~ln(i\hat{D}+i\hat{M}+\Delta)=Tr~ln(
i\hat{D}+i\hat{M})+Tr((i\hat{D}+iM)^{-1}\Delta)-
\ee
$$-\frac{1}{2}Tr[(i\hat{D}+iM)^{-1}\Delta (i\hat{D}+iM)^{-1}\Delta]\equiv
(-W^{(1)}+W^{(2)}+W^{(3)}).$$
where
\be
\Delta=i(e^{i\vec{\pi}\vec{\tau}}-1)\hat{M}_++
i(e^{-i\vec{\pi}\vec{\tau}}-1)\hat{M}_-,
\hat{M}=\hat{M}_++\hat{M}_-
\ee
The first term, $W^{(1)}$, contains no pions and describes
contribution of $q\bar{q}$ pairs interacting with background
field $B_{\mu}$, the quark being already massive due to
appearence of chiral mass $M$.

The second and third term, $W^{(2)}$ and $W^{(3)} $ are
depicted in Fig.2 (a) and (b) respectively.

 Expanding in (5.4) $\Delta$ in powers of  $\pi_i$ and
 keeping only quadratic terms  one obtains
 \be
 +W^{(2)}(\pi)+W^{(3)}(\pi) =
 \ee
 $$\int \frac{dk~dk'}{(2\pi)^8}
 \pi_a(k)\pi_a(k') N(k,k')$$
 where
 $$N(k,k') = \frac{1}{2}Tr[
 \frac{1}{i\hat{D}+i\hat{M}}i \tilde{M} (k+k') +
 \frac{1}{i\hat{D}+i\hat{M}}i \tilde{M} (k)
  \frac{1}{-i\hat{D}+i\hat{M}}i \tilde{M} (k') ]$$

and
$$<x|\tilde{M}(k)|y> = \frac{\varepsilon N}{2VN_c}\int
K(x,y,u)e^{iku}du~;~~\tilde{M}(0)=M~.$$

It is  easy to see in (5.6) and we show it explicitly in Appendix D  that
for $\pi_a=const$
i.e. for $k=k'=0$  the sum in the square brackets  on the r.h.s.  of (5.6)
vanishes,
signalling vanishing of the pion mass, as it should be for the spontaneous
CSB with zero current quark masses.

A similar analysis for other mesons can be done and results are
similar to those of [21]
. We note that masses of all mesons, other than $\pi_i$, do not
vanish and are of the order of typical hadron mass.

We turn now  to the final topic of this section; effective chiral Lagrangian
for pions. In this case  we integrate out all meson degrees of freedom exept
that of the pion, since pion  is the lightest particle dominating at small
momenta. From (5.4) we obtain
\be
W(\pi) = - Tr~ ln (i\hat{D}+i\hat{M}\hat{U}_5)
\ee
where
\be
\hat{U}_5 =U\frac{1-\gamma_5}{2} +U^+ \frac{1+\gamma_5}{2} =
e^{i\pi_i\tau_i\gamma_5}
\ee

The expression (5.7) reduces to the chiral Lagrangian obtained in [21,16]
when $\hat{M} \rightarrow M(0) $ and $B_{\mu} \rightarrow 0$, so that
$i\hat{D} \rightarrow i \hat{\partial}$.

In the form given in (5.7) the gauge invariance is seen explicitly (for that
the factor $\Phi(x,R,y)$ should be kept in $\hat{M}$ as is given in (3.15)).

The effective action (5.7) describes quark  pairs (integrated out)
propagating in the confining gluon field $B_{\mu}$ and the chiral field
$\hat{U}_5$ - the remaining degrees of freedom -- those of pions $(\pi_i)$
and gluons $(B_{\mu})$.

 We can compare (5.7) with the  standard chiral Lagrangian [2,3]
 \be
 W_{eff} (\pi) = \frac{F^2_{\pi}}{4} \int d^4 x Tr
 \Lambda_{\mu}\Lambda_{\mu}+
 \frac{N_c}{240\pi^2} \int d^5x \varepsilon _{\alpha p \gamma \delta
 \varepsilon} Tr \Lambda_{\alpha} \Lambda_{\beta}\Lambda_{\gamma}
 \Lambda_{\delta} \Lambda_{\varepsilon} + ...
 \ee
 where
  $$\Lambda_{\mu} = U^+ i\partial_{\mu} U.$$

Comparing (5.6) and (5.7) with (5.9) we  deduce the expression for the pion
decay constant $F_{\pi}$.
To this end one should average in $Z_{QCD}$ (4.7) over fields $B_{\mu}$,
which in the leading $N_c$ order results in averaging over $B_{\mu}$ the
effective action $W(\pi)$ (5.6--5.7). Since $N(k,k')$ includes trace over
coordinates, one has $<N(k,k')>_B\equiv(2\pi)^4\delta(k+k') N(k)$, and
$N(k)=k^2N_0(k)$. Finally comparing (5.6) to (5.9) we have
 \be
 \frac{1}{2} F^2_{\pi} =
N_0(k=0)
\ee
When $B_{\mu}$ is absent, we come back to Eq.(68) of [10]. For details see
Appendix D.

\section{Correlation functions and the double nature of pion}
\setcounter{equation}{0}
\def\theequation{6.\arabic{equation}}

Our starting point now is the QCD partition function where we keep quark and
 meson degrees of fredom (in addition to  $B_{\mu}$). To obtain it , we
 integrate (4.7) over $dg_+dg_-$, which effectively reduces to insertion
 into $\bar{W}_2$ in (4.7) $g_+=g_-=g_0$. We obtain
 \be
 W(L,R,\psi)= \int d^4xd^4x'\psi^+_f(x) [ i\hat{D}+i(1+\sigma+\eta)
 UV\frac{1-\gamma_5}{2}\hat{M}+
 \ee
 $$+ (1+\sigma-\eta)
 VU^+\frac{1+\gamma_5}{2}\hat{M}]_{fg,xx'}\psi_g(x')+
 \frac{N}{2V}\int d^4x (\sigma^2(x)+\eta^2(x)) $$
where $U,V$ are given in (5.2).

We now can calculate the $q\bar{q}$ correlation function
\be
\Pi^{\Gamma}(x,y) = < \psi^+(x)\Gamma\psi(x) \psi^+(y) \Gamma\psi(y)>
\ee
where $\Gamma=1, \gamma_5,\gamma_{\mu},\gamma_{\mu}\gamma_5,
\sigma_{\mu\nu}$. The result is
\be
\Pi^{\Gamma}(x,y) =\frac{1}{Z} \int D\Phi_i
D\mu(B)e^{-W(L,R)}\{Tr(S(x,y;\Phi,B)\Gamma S(y,x,\Phi,B)\Gamma)-
\ee
$$Tr~S(x,x;\Phi,B)
Tr~S(y,y;\Phi,B)\}$$
Here $W(L,R)$ is the effective action (5.3); $\Phi_i(x)~i=1,2,...8$ denotes
the set of meson variables; $\sigma(x),\eta(x), \sigma_a(x),
\pi_a(x)~.~~S(x,y;\Phi,B)$ is the quark propagator in the external field of
mesons $\{\Phi\}$ and confining gluons $\{B_{\mu}\}$.
\be
S(x,y;\Phi,B) = [ i\hat{D}+i(1+\sigma+\eta)
 UV\frac{1-\gamma_5}{2}\hat{M}+
 \ee
 $$+ i(1+\sigma-\eta)
 VU^+\frac{1+\gamma_5}{2}\hat{M}]^{-1}$$
 Two contributions in eq. (6.3) are shown in Fig.3a,b, we shall call them
 one (quark) loop and two--(quark)  loop respectively. This however is
 somewhat misleading since the action $W(L,R)$ corresponds to the
 superposition of any number of quark loops , since
 \be
 W(L,R)= -Tr~lnS^{-1}+\frac{N}{2V}\int d^4x (\sigma^2(x)+\eta^2(x))
 \ee

   Therefore our terminology refers only to the interior of the curly brackets
 in (6.3) with understanding that these graphs should be integrated with the
 weight shown in (6.3). Now we discuss the result of integration of diagrams
 (a) and (b) of Fig. 3 over bosonic fields in (6.3). To this end we expand
 $W(L,R) \equiv W(\Phi_i)$ in powers of $\Phi_i$ up to the quadratic terms
 [21].  \be W(\Phi_i)=\frac{N}{V} \int dx~dy~\Phi_i(x)
 {\cal{L}}_i(x,y)\Phi_i(y)
 \ee
  The term ${\cal{L}}_i$ gets
 contribution from two diagrams depicted in Fig.  2(a,b). For the case of
 pions ${\cal{L}}_{\pi}$ can be read off from (5.6).  For small values of
 momenta the Fourner transform ${\cal{L}}_i(k=0)$ plays the role of meson
 mass. As we discussed in section 5 contributions to ${\cal{L}}_{\pi}(k=0)$
 of Fig. 2(a) and (b) cancel each other establishing in this way the
 Goldstone theorem. One can easily see that ${\cal{L}}_i(x,y)$ do not
 depend on $N_c$ (since also $\hat{M}$ does not depend on $N_c$).  On
 the other hand, \be \frac{N}{V} = \frac{g^2}{8\pi^2}<F^2>_{inst}\sim 0(N_c)
 \ee

 Since $W(\Phi)\sim \frac{N}{V} \sim N_c$, the exchange of a meson yields
 correction $0(1/N_c)$ -- in other words one can tell that the coupling
 constant of a quark to the meson
 \be
 g_{\Phi q\bar{q}}\sim \frac{M}{F_{\phi}} \sim 0(N_c^{-1/2})
 \ee
 Hence the contribution of the diagram Fig.3(b) is $0(N_c^2\cdot
 \frac{1}{N_c})=0(N_c)$ just as that of the diagram Fig. 3(a). This is in
 contrast to the situation with gluon exchanges for the diagram of Fig.4,
 which contributes $0(N_c^2\cdot\frac{1}{N_c^2})=0(N_c^0)$ --  this is the
 so-called OZI supressed diagram, just as the diagram of Fig. 3(b). However
 in the vector and tensor channels the latter does not exist  (tensor mesons
 appear in the $1/N_c$ corrections, see [21,24]) and therefore in these
 channels OZI suppression can be explained by the smallness of the diagram
 Fig.4. At the same time, in the scalar and pseudoscalar channels there is
 no suppression since the diagram of Fig.3(b) provides for the  amplitude
 roughly the same amount as the OZI allowed diagram of Fig. 3(a).
 This  conclusion of no OZI suppression in scalar and pseudoscalar channels
 has been obtained before in [29] using another arguments.

 We now turn to the important question  about the double role of the pion --
 as the Goldstone particle and as the $q\bar{q}$ bound state. Our present
 formalism is a convenient setting for the study of this question and we
 shall actually have in mind all scalar and pseudoscalar particles (to be
 compared later with vector and tensor particles).

 Let us start with the effective action (6.1). It contains both quark and
 meson degrees of freedom, but here they enter not on equal footing. Quarks
 $\psi_f,\psi_f^+$ are supplied with kinetic term and are respectable
 quantum field--theoretical quantities. In contrast to that meson fields
 $\Phi_i$ enter as auxiliary fields, they do not possess kinetic terms and
 should be considered as external fields to be averaged out with the given
 weight.  When one integrates out  quarks, one obtains an effective chiral
 Lagrangian as in (5.7--5.8), where the chiral pion field now inherits the
 kinetic term and the full QFT status. But in doing that (integrating out
 quarks) we average over all quark structure of the pion and
 "see only its chiral face".  Quantitatively both faces of the pion
 are given by (6.3), or diagrams Fig.3(a,b). While the diagram
 of Fig.3(a) is the usual one considered in quark models, the diagram
 Fig.3(b) is pertinent to the chiral degrees of freedom -- it contains the
 chiral propagator $({\cal{L}}_i)^{-1}_{xy}$.

 Now in the limit of large $N_c$ the diagram Fig. 3(a) will contain only
 gluon $(B_{\mu})$ exchanges and provides only poles at $p^2=m^2_a(n),
 n=1,2...$.
 The chiral propagator $({\cal{L}}_i)^{-1}_{xy}$ provides a pole at
 $p^2=m^2_b$; for pion $m^2_b=0$. How these two types of poles coexist?

 For pion when $p^2$ is small, $p^2\ll R_0^{-2}, \rho^{-2}$ the situation is
 relatively simple. In this region only the  diagram Fig. 3(b) has the pole
 and it is dominant. When $p^2$ increases and is close to the pole $m^2_a$,
 the same pole appears in ${\cal{L}}_i(p)$.
 This is the signal that the expansion in powers of $\Phi_i$ done in (6.6)
 and considering the diagrams of Fig. 3(b) is meaningless -- the notion of
 the meson mass is not useful when it is  strongly $p$ -- dependent and even
 has a pole at $p^2=m^2_a$. Formally, the diagrams Fig.3(b) acquire a pole
 at $m^2_a$, so that the sum of contributions of Fig.3a and Fig.3b according
 to (6.3) can be written as
\be
\pi_5(k) = \pi_5^{(0)}(k) - \pi_5^{(0)}(k) \frac{M^2}{N(k)}\pi_5^{(0)}(k)
\ee
where $\pi_5^{(0)}(k)$ is the contribution of the quark loop of Fig.3a.

Now $N(k)$ can be written as (in the limit $\rho\rightarrow 0$)
\be
N(k) =C+M^2(0)\pi_5^{(0)}(k)
\ee
The "quark model poles" appear in $\pi_5^{(0)}(k)~,$
\be
\pi_5^{(0)}(k)=
 \frac{\lambda^2}{k^2+m_a^2}
 \ee
 One can see in (6.9) (at least in the limit $\rho\rightarrow 0$, i.e. for
 $k\rho \ll 1$) that "quark model poles" (6.11) exactly cancel, and only
 Nambu-Goldstone pole at $k=0$ due to $N(k\rightarrow
 0)=\frac{1}{2}F^2_{\pi}\cdot k^2$ survives. Thus the Goldstone theorem
 manifests itself in our case, as it should be since chiral symmetry breaks
 spontaneously.

  \section{Conclusions and outlook}

  We have shown that the QCD vacuum with confinement and topological charges
   can be adequately described by the model, in which confining
 configurations are added to instantons.

 On one hand, instantons are stabilized by  confinement, and become
 more dilute, so that they can be properly treated in the instanton
 gas approximation.

 On  an other hand, quarks and gluons are confined in the model, and
 unphysical features of the instanton gas or liquid model, where quarks
 propagate freely, are now absent.

 Therefore in this paper we have obtained a realistic approach, in which
 quarks, gluons and Goldstone bosons can be treated simultaneously on the
 fundamental level. The basic effective action is given in (5.3) and can be
 used for diagrammatic expansion in powers of Goldstone exchanges or
 integrating over all bosonic fields, in particular when there is an
 extremum corresponding to a selfconsistent bosonic field  of solitonic type.

 The latter might be important for baryons.

 In the limit when confining configurations vanish we come back to the
 effective chiral Lagrangian obtained earlier [21],[30].

 The framework suggested in the paper can be used for the calculations of
 all effects where chiral physics and confinement are both important. A
 systematic quantitative study is planned in subsequent publications.

 \newpage \begin{center} {\bf Figure captions} \end{center}

 Fig. 1. Graphical representation of the heavy--light Green's function,
 eq.(3.17), The straight line from $x$ to $y$ corresponds to the
 heavy--quark path, the light--quark path is typically away at a distance of
 $R_c\sim 1 fm$ from the heavy quark.

 The chiral mass insertion (second term in Eq. (3.21)) is shown near the
 instanton position $R_i$ with nonlocality of the order of $\rho\sim 0.2~
 fm$.

 Fig. 2. Graphical representation of two terms in the effective  action
 (5.4). Part (a) corresponds to $W^{(2)}$ and part (b) to $W^{(3)}$, solid
 lines denote quark propagator $(iD+iM)^{-1}$, broken lines -- emitted
 pion field $\pi_i$ from expansion of $\Delta$, Eq. (5.5).

 Fig. 3.  Graphical representation of the $q\bar{q}$ correlator
 $\Pi^{\Gamma}(x,y)$, Eq. (6.3). Fig. 3a corresponds to the "one--loop
 term" (first term inside the curly brackets of (6.3)), while Fig. 3b
 corresponds to the "two--loop term" (second term inside  the curly
 brackets). Broken line denotes the boson propagator which appears when one
 expands $S(x,x,\Phi,B)$ in powers of $\Phi$.

    Fig. 4. The OZI violating two--gluon exchange diagram obtained from
 expansion in $B_{\mu}$ of the two--loop term of Eq. (6.3).

\newpage

 \underline{{\bf Appendix A}}
\setcounter{equation}{0}
\def\theequation{A.\arabic{equation}}
 \begin{center}
 {\bf Quark Green's function in the instanton field}
 \end{center}

 In this appendix we remind the reader the expansion of the quark Green's
 function $g$ in the pure instanton field $A^{(i)}(x)$ in the unitary gauge,
 which was found in [31].
 \be
 G(x,y) = (i\hat{D}-im)\Delta \frac{1+\gamma_5}{2} + i\Delta\hat{D}\cdot
 \frac{1-\gamma_5}{2}+\frac{1}{im} u(x,y)
 \ee
 where $u(x,y)$ is the contribution of zero modes
 \be
 u(x,y)=u_0(x)u_0^+(y)
 \ee
 while $\Delta(x,y)$ is the Green's function of scalar particles in the
 instantonic field in the unitary gauge
 \be
 (-D^2_{\mu} + m^2) \Delta (x,y) = \delta^{(4)} (x-y)
 \ee
 One can obtain expansion of $\Delta$ in powers of $m$ [32,29] in the
 singular gauge \be \Delta(x,y) = \phi^{-1/2}(x) \tilde{\Delta}(x,y)
 \phi^{-1/2}(y)~,~~ \phi= 1+\frac{\rho^2}{x^2} \ee \be \tilde{\Delta}(x,y) =
 \frac{1}{4\pi^2}
 \{\frac{1}{(x-y)^2}+\rho^2
 \frac{(\tau_-x)(\tau_+y)}{x^2(x-y)^2y^2}\}+0(m^2)
 \ee
 Consequently one has for $G(x,y)$, $x\sim y\sim \rho$
 \be
 G(x,y) = \frac{1}{im} u(x,y) (1+0(m\rho)
 \ee
  This can be compared with the spectral decomposition (2.4),
 with the result that the contribution of the nonzero modes is
 finite  for $m\rightarrow 0$.

 Now we can see the physical parameter of expansion in the
 ansatz (2.9) used here and earlier papers [10,21--24].

 One can state that the omitted terms in (2.9) are of the order $m\rho$
 (since $mx$ and $my$ which can also appear effectively enter our
 expressions for quark propagator or effective action at distances $x,y \leq
 \rho$).

 Thus $m\rho\ll 1$ is also accuracy of our approximation in (3.1). One can
 see that for a typical instanton size $\rho\approx 0.2 fm$, $u,d$ and $s$
 quarks satisfy condition $m\rho\ll 1$, while for $c$ quark this is already
 violated and all effective actions considered below are not applicable for
 the $c$ quark.
  \newpage

 \underline{{\bf Appendix B}}
\setcounter{equation}{0}
\def\theequation{B.\arabic{equation}}
 \begin{center}
 {\bf Shift of eigenvalues due to background $B_{\mu}$}
  \end{center}

 We study eigenvalues in the field of one instanton $A_{\mu}$ plus
   background $B_{\mu}$, which are to be found from the equation [12]:
   \be
   [-(\partial_{\mu}-ig(A_{\mu}+B_{\mu}))^2 - g\vec{\sigma}(\vec{E}(A+B)+
   \vec{B}(A+B))]\varphi_n=\lambda_n^2\varphi_n
   \ee
   where $A^a_{\mu} = \frac{2}{g} \eta_{a\mu\nu}\frac{x_{\nu}}{x^2+\rho^2}$
    $\frac{\rho^2}{x^2}$
    while $\vec{E},\vec{B}$ are colorelectric and colormagnetic
    fields respectively.

    Let us concentrate on the shift of the zero eigenvalue which we shall
    evaluate by perturbation theory
    \be
    \delta \lambda^2 =(\varphi_0\delta V\varphi_0)
    \ee
    where $\delta V$ is to be read off from eq. (B.1) and is due to
    $B_{\mu}\not= 0$. One can extract the phase factor out $\varphi_n$, as
    is done in (2.23)   and the rest is equivalent to the wave function in
    the Fock-Schwinger gauge for $B_{\mu}$;
    \be
    B_{\mu}(x)=\int^x_0
    \alpha(u)du_{\nu}F_{\nu\mu}(u)~,~~\alpha(u)=\frac{u}{x}
    \ee
    In $\delta V$ there are three terms:
    (i) linear in $B_{\mu}$ (ii) quadratic in $B_{\mu}~,~~g^2 B^2_{\mu}$
    (iii) proportional to $\vec{E}+\vec{B}$. Due to symmetry reasons only
    quadratic term contributes to $\delta\lambda^2$.

    We have
    \be
    \delta \lambda^2= 2g^2\int^{\infty}_0\frac{r^3dr\rho^2
    B^2_{\mu}}{(\rho^2+r^2)^3}=
    \ee
    $$
    =2g^2\int^{\infty}_0\frac{y^3dy}{(1+y^2)^3} \int^{\rho y}_0
    \alpha(u) \alpha(u')du_{\nu}du'_{\nu}<F_{\nu\mu}(u)F_{\nu'\mu} (u')>$$
    Here we have introduced for an estimate the average value of
    $<F(u)F(u')>$ which at $u=u'$ should be less or equal to the gluonic
    condensate of [33]:
    \be
    \frac{\alpha_s}{\pi}<F^a_{\mu\nu}F^a_{\mu\nu}>= 0.012 GeV^4
    \ee

    Taking into account that $<F(u)F(u')>$ falls off at distances
    $|u-u'|\sim T_g \approx 0.2 fm$ [34], one can compute the integrals in
    (B.4) to obtain
    \be
    \delta \lambda^2 \cong \frac{\rho^2}{8} \frac{\alpha_s}{\pi}<F^aF^a>\leq
    0.0015 Gev^4
    \ee
    In getting upper bound on the r.h.s.  of (B.6) we use the fact that part
    of gluonic condensate should be due to instantons and in our case only
    that part of $<F^aF^a>$ enters in (B.6) which  is due to confining
    background $B_{\mu}$.
     Hence we have average shift of the zero eigenvalue $<|\delta
     \lambda|>\leq 40 MeV$.  This is small as compared to $\rho^{-1}
     \approx 1 Gev$ , and therefore $|\delta\lambda|\rho \ll 1$.
      \newpage

 \underline{{\bf Appendix C}}
\setcounter{equation}{0}
\def\theequation{C.\arabic{equation}}
 \begin{center}
 {\bf Derivation of the effective action (3.1)}
 \end{center}

 We start with the  general form of the partion function for the quarks of
 flavour $f=1,... N_f$ in the field (2.1)
 \be
 Z=const \int D{\mu}(B) D\psi D\psi^+ exp - \int\psi^+_f S^{-1}\psi_f dx
 \ee
 where the action of gluonic field is  included in $D{\mu}(B)$.

 We now transform $S^{-1}$ in (C.1) to make explicit contribution of zero
 modes
 \be
 S^{-1}\equiv -i\hat{\partial}-g\hat{B}-g\sum\hat{A}^{(i)}-im
 =S_0^{-1}+\sum^N_{i=1}[(S^{(i)})^{-1}-S_0^{-1}]
 \ee
 Using (2.5) and (2.8) after some algebra one obtains
\be
S^{-1}(x,y) = S^{-1}_0(x)\delta (x-y)
   +\sum^N_{i=1}\sum^K_{n,q=0}S^{-1}_0(x)u^{(i)}_n(x)
   (im-\hat{\lambda}-\hat{V})^{-1}_{ii,nq}u^{(i)+}_q(y)S^{-1}_0(y)
   \ee
   where matrix elements of $\hat{V}$ are
   \be
   V_{nq,ii} =\int u^{(i)+}_n(z)(-i\hat{D}(B)-im)u_q^{(i)}(z)dz
   \ee
   It is easy to check that inverion of $S^{-1}$ given by (C.3) yields
   (2.10).

    For $n=q=0~~u^{(i)}_0$ has a definite chirality and therefore $V_{00
    ,ii}=-im$. Hence, keeping only zero models in (C.3), as it is done
    e.g.in [21--24], one obtains a term $0(\frac{1}{2im-\lambda})$ in
    $S^{-1}$ in (C.3) which was not present from the beginning in (C.2)

     The reason for this apparent paradox can be identified as an improper
     omission of terms with $n,q\not= 0$. Indeed, matrix elements
     $V_{11,ii}$ and $V_{10,ii}$ are of the order of $0(\frac{1}{\rho})$,
     since eigensolutions of (2.3) with $n\not= 0$ have no definite
     chirality and have nonzero matrix elements even in the
     limit $m=0$.

     Therefore $(im-\hat{\lambda}-\hat{V})^{-1}_{nq} \sim
     (det(im-\hat{\lambda}-\hat{V}))^{-1}$ is finite for $m\rightarrow 0,
     \lambda_0 \rightarrow 0$, and this fact resolves the apparent paradox
     occurring in derivation in [21, 24].

      Insertion of (C.3) into (C.1) yields
      \be
      Z=const\int D\mu (B)D\psi D\psi^+exp(-\int\psi^+_fS^{-1}_0\psi_fdx)
      exp \int{\cal{L}} dxdy
      \ee
      where
\be
{\cal{L}}=-\sum_f \psi^+_f(x)
   \sum^N_{i=1}\sum_{n,q}S^{-1}_0(x)u^{(i)}_n(x)
   (im-\hat{\lambda}-\hat{V})^{-1}_{ii,nq}u^{(i)+}_q(y)S^{-1}_0(y)\psi_f(y)
   \ee
One can notice that $\int{\cal{L}}dx~dy$ contains fermionic operators of the
       type:
\be
\int\psi^+_f(x)S^{-1}_0(x) u^{(i)}_n(x)dx \equiv \Psi^{(i)}_n(f)
\ee

Due to anticommutativity of $\psi^+_f$ the product
       $\Psi^{(i)}_n(f)\Psi^{(i)}_n(f)$ vanishes. Therefore in expansion of
       $exp {\cal{L}}$ only finite number of terms survives. Namely, if we
       for simplicity keep only zero modes, $n=q=0$ in (C.6) then one can
       write.
       \be
       exp \int {\cal{L}} dxdy= \prod^{N,N_f}_{i,f}
       (1-\Psi_0^{(i)}(f) (im-\hat{\lambda}_0-\hat{V})^{-1}\Psi_0^{(i)+}(f))
       \ee

       This form coincides with that obtained in [21--24]. Our derivation
       which has used (C.3) is more direct than prosented in [24]. Eq. (C.8)
       also coincides with the form originally suggested in [21] up to a
       change $im_f\rightarrow (im-\lambda_0-\hat{V})^{-1}_{00}$. This
       overall factor can be taken out to redefine a normalization of the
       partition function and one comes to (3.1).

        One can also show that nonzero modes can be neglected since the
        ratio
        $(im-\hat{\lambda}_n-\hat{V})^{-1}_{nn}
        /(im-\lambda_0-\hat{V})^{-1}_{00}$ is of the order  $0(m\rho)$.
         Namely for the two-channel situation,
         $n=0,1$ one has
         \be
         (im-\hat{\lambda}-\hat{V})^{-1}_{00}
         =\frac{im-\lambda_1-V_{11}}{det \Delta}~,~~
        (im-\hat{\lambda}-\hat{V})^{-1}_{11}=
        \frac{2im-\lambda_0}{\det\Delta}
        \ee
        Now take into account that $\lambda_1\sim V_{11}\sim\rho^{-1}$,
        while $m$ and $\lambda_0$ are much smaller (see Appendix B).

\newpage

 \underline{{\bf Appendix D}}
\setcounter{equation}{0}
\def\theequation{D.\arabic{equation}}
 \begin{center}
 {\bf Study of the pion inverse propagator, $N(k,k')$ in (5.6)
 } \end{center}

 We first prove that $N(k=k'=0)=0$. From (5.6) one has
 \be
 N(0,0) = \frac{1}{2} Tr[\frac{1}{i\hat{D}+iM}iM+\frac{1}{i\hat{D}+iM}
 M\frac{1}{-i\hat{D}+iM}M]=
 \ee
 $$=\frac{1}{2} Tr[\frac{1}{i\hat{D}+iM}M\frac{1}{-i\hat{D}+iM}\hat{D}]=$$
 $$=\frac{1}{4i}
 Tr(\frac{1}{-i\hat{D}+iM}\hat{D}+\frac{1}{i\hat{D}+iM}\hat{D})$$

 Consider the tranformation of inversion $P$ of all coordinates,
 $x_{\mu}\rightarrow - x_{\mu}~,~~B_{\mu} \rightarrow  -B_{\mu}$.

 From (D.1) one can see that under $P$ transformation $N(0,0)$ changes sign.
 Therefore, when $N(0,0)$ is integrated over all gluonic fields and is a
 number, it should vanish
 \be
 <N(0,0)>_B=0
 \ee

 From the invariance properties of the vacuum with respect to the shift of
 coordinates, one can deduce that $<N(k,k')>_B$ written as
 \be
 <N(k,k')>_B= \int du~du'e^{iku+ik'u'}<\frac{1}{2} Tr
 [\frac{1}{i\hat{D}+iM}im(u)\delta (u-u')+
 \ee
 $$+\frac{1}{i\hat{D}+im}m(u)\frac{1}{-i\hat{D}+iM}m(u')]>_B
 $$
 where
 \be
 <x|m(u)|y>=\frac{\varepsilon N}{2VN_c}K(x,y,u),
 \ee
 has the property
 \be
 <N(k,k')>_B=(2\pi)^4\delta (k+k')N(k).
 \ee

 Expanding now $N(k)$ and again using invariance of the vacuum we get
 \be
 N(k)= k^2N_0(k).
 \ee
 Finnaly we report in this Appendix the calculation of $<N(k,k')>_B$ in the
 limit $\rho\rightarrow 0$. We have for the first term in (5.6)
 \be
 N^{(1)}= <\frac{1}{2} tr \int dx~dy~ du
 \ee
 $$ \cdot e^{i(k+k')u}
 S(x,y,B)\frac{\varepsilon N}{2VN_c} i\hat{D}\bar{\varphi}(y-u)
 \Phi(y,u,x)\bar{\varphi}^+(x-u)i\hat{D}>_B
$$

  When $\rho\rightarrow 0$, both $x$ and $y$ tend to $u$ and one can replace
 $S(x,y;B)\rightarrow S(u,u;B)$. Integration over
 $D(x-u)$ and $(y-u)$ yields
 \be
 N^{(1)} =(2\pi)^4\delta(k+k')const
 \ee

 Analysis of the second term on the r.h.s. of (5.6) yields in the limit
 $\rho \rightarrow 0$
 \be
 N^{(2)} = \int du~dw~e^{iku+ik'w}<\frac{1}{2}
 tr~S(w,u;B)M(0)S(u,w;B)M(0)>_B
 \ee

 Finally we can rewrite the sum $N^{(1)}+N{(2)}$ as
 \be
 <N(k,k')>_B=N^{(1)}+N^{(2)}=
 \ee
 $$=(2\pi)^4\delta(k+k')const+M^2(0)\pi_5(k))
 $$
 This justifies eq. (6.10) used in the text.

\newpage
\begin{center}
{\bf ACKNOWLEDGEMENTS}
\end{center}

This work has been started while the author was at the Institute of
     Theoretical Physics of the Heidelberg University as an awardee of the
     Alexander von Humboldt -- Stiftung. It is a pleasure for him  to thank
     the Alexander von Humboldt Stiftung for a financial
     support and the Institute and in particular Professor
     H.G.Dosch for a cordial hospitality.

     Useful discussions with B.L.Ioffe and members of the ITEP
     theoretical seminar are gratefully appreciated.

     \end{document}